\documentclass[12pt]{elsart}
\usepackage{epsfig}
\date{1 October 2004}
\begin{document}

\begin{frontmatter}
 
\title{Gas  gain on single wire  chambers filled with  pure isobutane  at low
 pressure}   

\thanks[contact]{Corresponding    author.     \\    E-mail    address:
davydov@triumf.ca                    (Yuri                   Davydov)}
\author[a,b]{Yu.I.~Davydov\thanksref{contact}},
\author[a]{R.~Openshaw},                      \author[b]{V.~Selivanov},
\author[a]{G.~Sheffer}   \address[a]{TRIUMF,   4004   Wesbrook   Mall,
Vancouver, BC, Canada  V6T 2A3} \address[b]{RRC "Kurchatov Institute",
Kurchatov sq. 1, Moscow, 123182, Russia}

\begin{abstract}
The gas gain of single-wire  chambers filled with isobutane, with cell
cross-section  12x12~mm   and  wire  diameters  of  15,   25,  50  and
100~$\mu$m,  has  been   measured  at  pressures  ranging  12-92~Torr.
Contrary  to  the experience  at  atmospheric  pressure,  at very  low
pressures the  gas gain  on thick  wires is higher  than that  on thin
wires at the same applied high voltage as was shown in~\cite{Davydov}.
Bigger wire  diameters should  be used in  wire chambers  operating at
very low pressure if multiple scattering on wires is not an issue.
\end{abstract}


\end{frontmatter}
 
\section{Introduction}

Electric field in the cylindrical wire chamber is defined as

\begin{equation}\label{eqn:efiled}
E=\frac{V}{r ln(b/a)}      
\end{equation}

\noindent where  {\it a} and {\it  b} are the wire  and cathode radii,
and {\it r} is the distance from the wire center. Electric field drops
very fast with distance from the wire surface and at normal conditions
most  gas  amplification  takes  place  within  3-5  wire  radii.   At
atmospheric pressure  small diameter wire requires  much lower applied
high voltage in order to reach the same gas gain as on larger diameter
wire. This  is the reason  why small diameter  wires are used  in wire
chambers.  The  situation changes when  gas pressure decreases  to the
level of a few tens of Torr.  As was recently shown~\cite{Davydov}, in
chambers with the same geometry the gas gain becomes higher on a thick
wire  compared  to that  on  a  thin wire  at  the  same applied  high
voltages. \\
\indent The  present work was  done to verify  the gas gain  on single
wire chambers as a function of wire diameters and gas pressure.

\section{Experimental setup}

\indent  Tests  were  carried  out  with  single-wire  chambers.   The
chambers have a cell cross section 12x12~mm and a wire length of about
20~cm.  Chambers  are  made   of  aluminum  alloy  with  double  sided
aluminized mylar on two sides serving as cathodes.  Chambers with wire
diameters 15,  25, 50 and  100~$\mu$m have been tested.   The chambers
were placed  in an aluminum  vacuum box.  The  box was pumped  out and
filled with  pure $iso-C_{4}H_{10}$  a few times,  and finally  it was
pumped out  to the  required pressure for  the test  measurements. The
measurements  have  been   done  at  pressures  of  92,   52,  32  and
12~Torr.  The pressure  was monitored  with  a pressure  gauge with  a
precision of $\pm$1~Torr.  Chambers  were irradiated with an $^{55}$Fe
x-ray source collimated with a  1.6~mm thick G10 plate with 3~mm hole.
This plate  was placed directly over  the mylar cathode  with the hole
close to the cell edge.  Signals from the chambers were self triggered
and fed into a LeCroy 2249W ADC, with a gate width of 1~$\mu$s for all
tests.

\section{Results and discussion}
 
\indent $^{55}$Fe  x-ray photons have energy 5.9~{\it  keV} (80\%) and
6.49~{\it  keV}~(20\%).  The  x-rays undergo  only  photoabsorbtion in
pure $iso-C_{4}H_{10}$ at these energies.  The released electrons have
a  range   of  about   700-750$\mu$m  in  pure   $iso-C_{4}H_{10}$  at
atmospheric pressure.  At 92~Torr  the range becomes about 6~mm, while
it is about 11~mm, 17~mm and 45~mm at 52, 32 and 12~Torr respectively.
At low pressure, there is a high probability that electrons will leave
the  12x12~mm cell before  they lose  all their  energy.  Some  of the
electrons lose  all their energy  within the cell,  but do not  give a
full avalanche if the ionization electrons originate close to the wire
surface.  As a  result, the measured charge spectrum  has a continuous
distribution with a  full photoabsorption peak at the  high energy end
from  the electrons  which stop  inside of  the cell  and give  a full
avalanche. \\
\indent  The  total x-ray  photoabsorption  results  in 5.9~{\it  keV}
energy loss inside of a  cell.  The average energy required to produce
an   ion-electron   pair   in   pure  $iso-C_{4}H_{10}$   is   23~{\it
eV}~\cite{Sauli}.   Thus, the photoabsorption  of 5.9~{\it  keV} x-ray
photons    results   in    about      256    electrons   in    pure
$iso-C_{4}H_{10}$. This number will be  used to calculate the gas gain
in the test chambers. \\
\begin{figure}[t]
\centering
\begin{minipage}[c]{0.8\textwidth}
\vspace{5mm}
\begin{center}
\epsfig{file=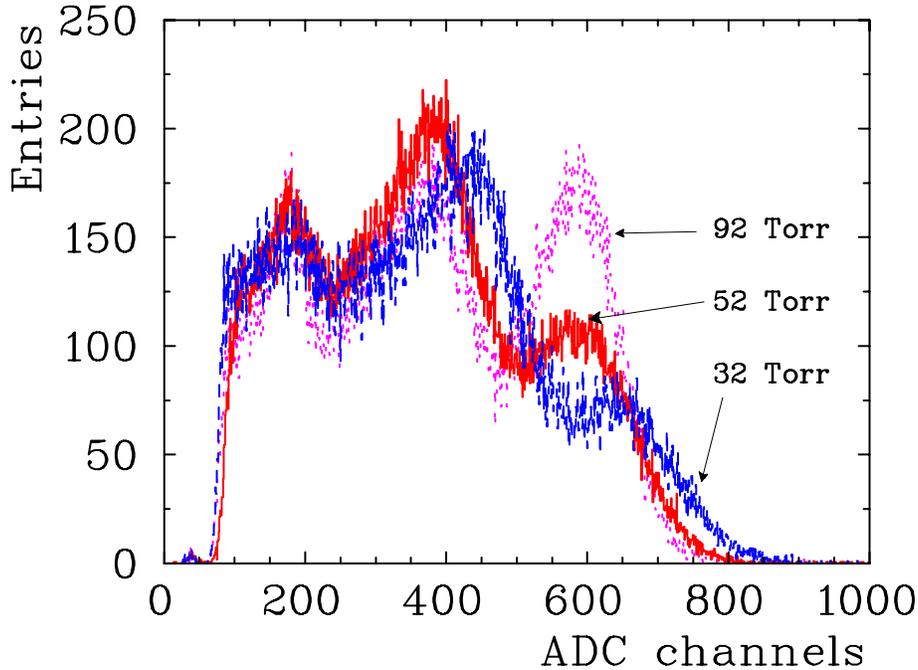,width=12cm,angle=0}
\end{center}
\caption{Charge spectra  on the single-wire chamber,  with a 25~$\mu$m
diameter wire, filled with pure $iso-C_4H_{10}$ at pressure 92, 52 and
32   Torr.    Applied   voltages   are  1150~V,   1000~V   and   950~V
respectively.Chamber irradiated with  $^{55}$Fe x-rays.  Gas gains are
in  the  range  9000-13500.   The  total  absorption  peak  resolution
degrades with decreasing gas pressure.}
\label{spectrum_pressure}
%\vspace{10mm}
\end{minipage}\hfill
\end{figure}
\indent  Figure~\ref{spectrum_pressure} presents  charge distributions
measured on  the 25~$\mu$m  wire at pressures  of 92, 52  and 32~Torr.
Spectra  were taken  at applied  high voltages  of 1150~V,  1000~V and
950~V respectively.  Gas gains  at these applied  voltages are  in the
range 9000-13500.   As mentioned  earlier the spectra  have continuous
distributions  with  full  photoabsorption  peaks  at  the  end.   The
fraction of  events in the  photoabsorption peak, and  its resolution,
drop with decreasing pressure.  All charge distributions were fit with
an  exponential function  to describe  the tail  of  continuous charge
distribution combined  with a  gaussian for the  photoabsorption peak.
The gaussian  function gives the  photoabsorption peak center  and its
sigma. The results of the gaussian  fit were used to calculate the gas
gain. \\
\begin{figure}[h]
\centering
\begin{minipage}[c]{0.8\textwidth}
\vspace{5mm}
\begin{center}
\epsfig{file=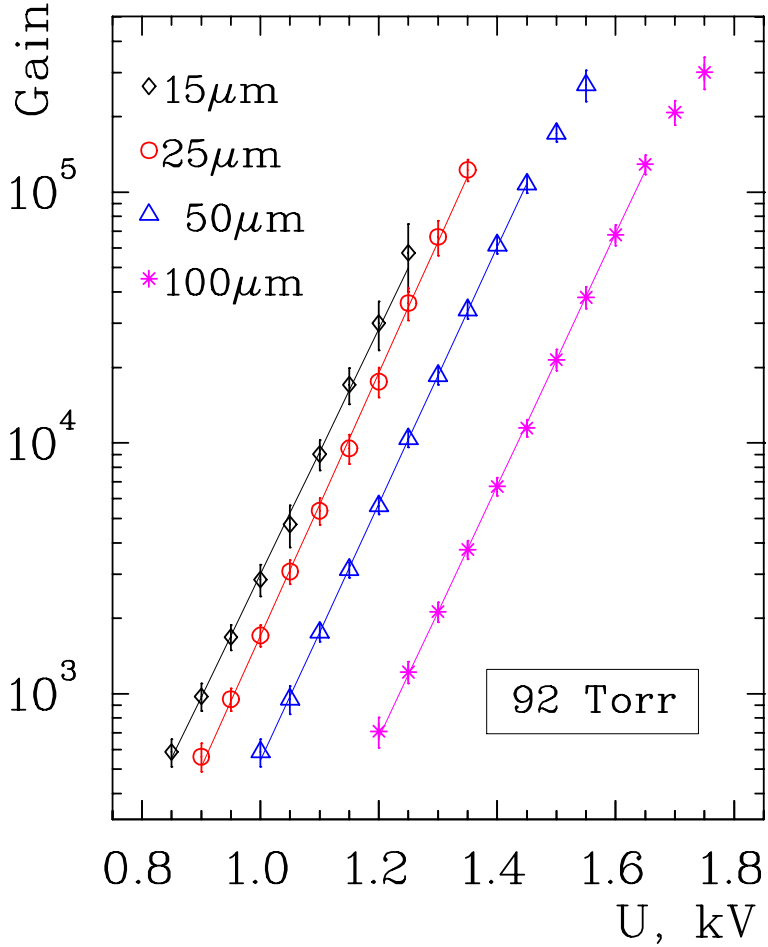,width=10cm,angle=0}
\end{center}
\caption{Gas gain versus high voltage of single-wire chambers with 15,
25, 50 and 100~$\mu$m diameter  anode wires.  Chambers are filled with
pure  $iso-C_4H_{10}$   at  92~Torr  and   irradiated  with  $^{55}$Fe
x-rays. The straight lines  represent exponential fits of experimental
points.}
\label{gain_92torr}
\vspace{5mm}
\end{minipage}\hfill
\end{figure}
\indent Figure~\ref{gain_92torr}  shows the gas gain as  a function of
applied voltage on all four tested wires at a gas pressure of 92~Torr.
Here and  on the next two  figures the straight  lines are exponential
fits of experimental points.  One  can see that at 92~Torr the thinner
wires   have   higher   gas    gain   at   the   same   applied   high
voltage. Resolution is  much poorer on the thin  wires, which resulted
in fewer points on the 15~$\mu$m and 25~$\mu$m wires. \\
\begin{figure}[t]
\centering
\begin{minipage}[c]{0.8\textwidth}
\vspace{5mm}
\begin{center}
\epsfig{file=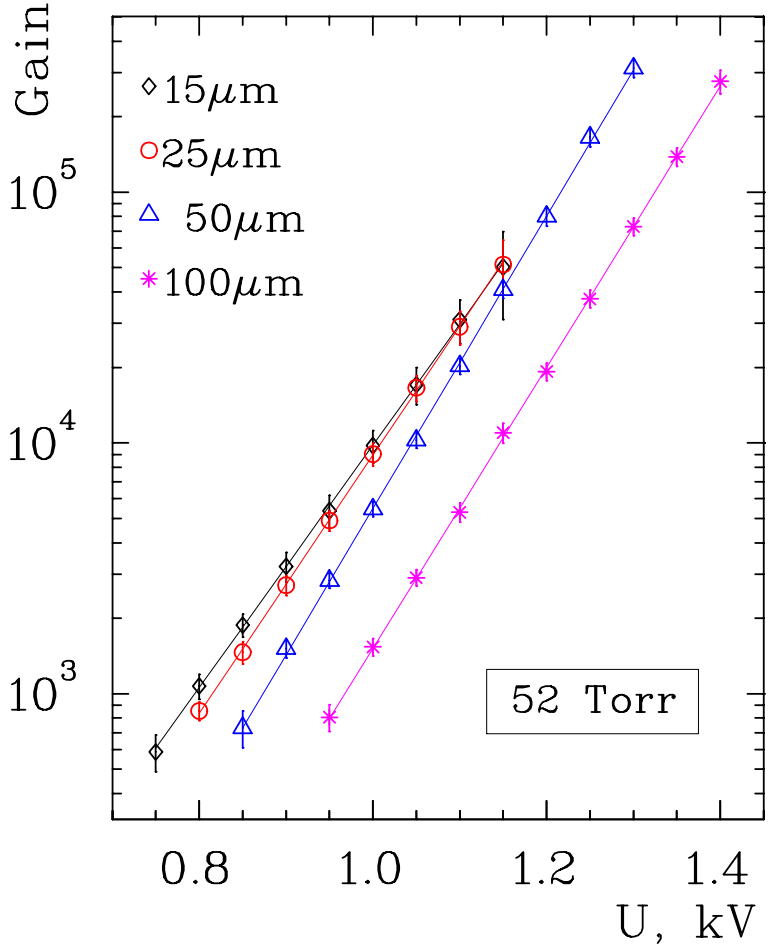,width=10cm,angle=0}
\end{center}
\caption{Gas  gain versus  high voltage  of  the same  chambers as  in
fig.\ref{gain_92torr}  at 52  Torr  pressure. Note  that  gas gain  of
15~$\mu$m wire is nearly the same as for 25~$\mu$m wire at \~1050~V.}
\label{gain_52torr}
\vspace{5mm}
\end{minipage}\hfill
\end{figure}
\indent At  a pressure  of 52~Torr (fig.~\ref{gain_52torr})  there are
visible changes compared with  92~Torr.  The smaller the wire diameter
the smaller the  slope of gas gain versus  high voltage dependence. At
lower voltages, the gain is higher on the 15~$\mu$m wire compared with
that  on the  25~$\mu$m wire.   Above  1000~V, gain  on the  15~$\mu$m
becomes  equal or  even  lower than  on  the 25~$\mu$m  wire and  even
approaches that on the 50~$\mu$m wire. \\
\begin{figure}[t]
\centering
\begin{minipage}[c]{0.8\textwidth}
\vspace{5mm}
\begin{center}
\epsfig{file=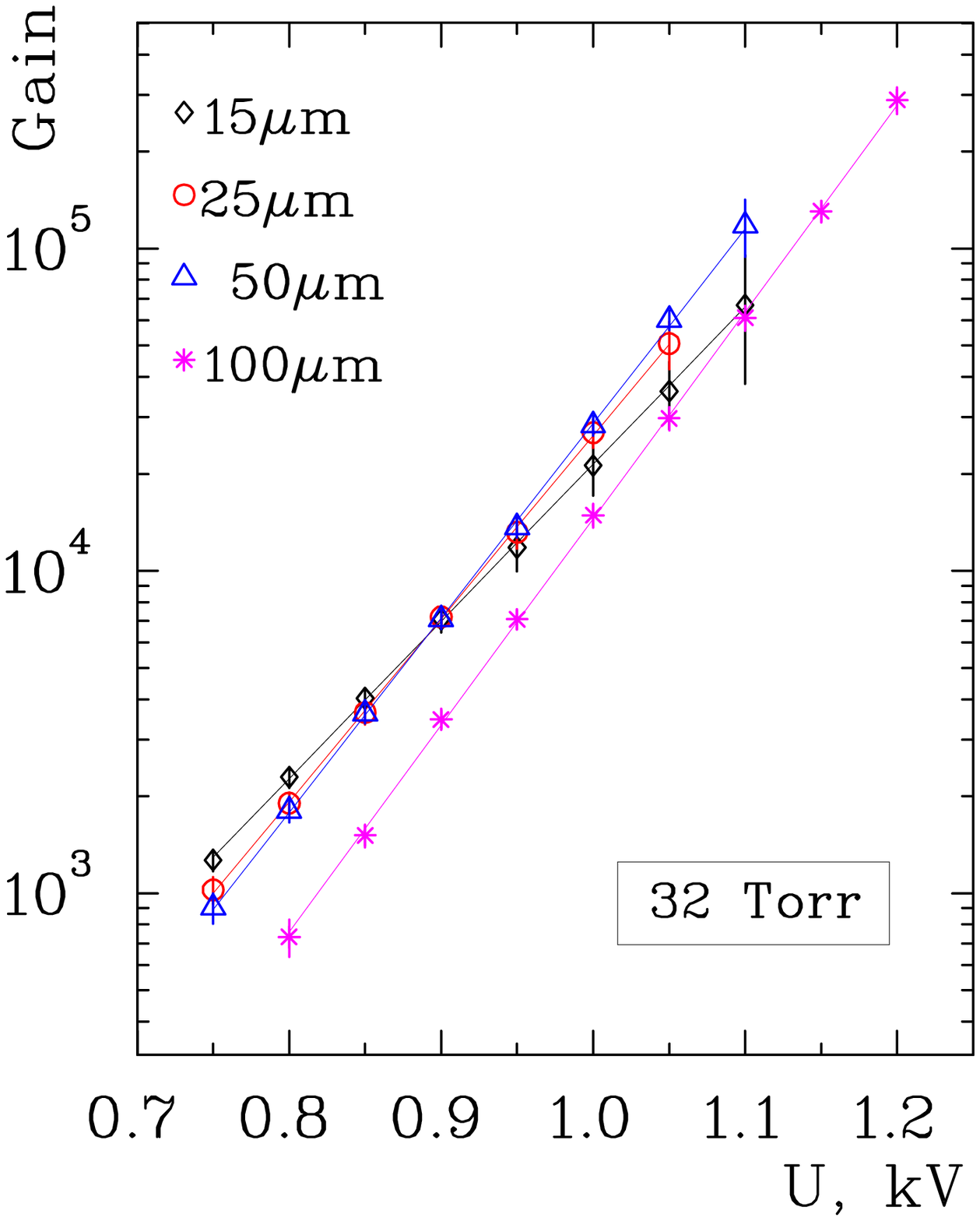,width=10cm,angle=0}
\end{center}
\caption{Gas  gain versus  high voltage  of  the same  chambers as  in
fig.\ref{gain_92torr} at  32~Torr pressure. At lower  voltages the gas
gain is higher on the thinner  wires at the same applied high voltage.
Eventually the order reverses for the three thinnest wires.}
\label{gain_32torr}
\vspace{5mm}
\end{minipage}\hfill
\end{figure}
\indent Figure~\ref{gain_32torr} depicts gas gains on the tested wires
at  a  pressure  of   32~Torr.  Resolution  becomes  very  poor  here,
especially for  the 15  and 25~$\mu$m wires.   As in the  two previous
cases, the  gas gain is higher  on thinner wires at  low voltages, but
that changes very quickly. At 900~V  gains on the 15, 25 and 50~$\mu$m
are  already equal  and  eventually reverse  the  order compared  with
higher pressure.  Above 1050~V the  gain on the 100~$\mu$m wire almost
reaches that on the 15~$\mu$m wire.\\
\begin{figure}[t]
\centering
\begin{minipage}[c]{0.8\textwidth}
\vspace{5mm}
\begin{center}
\epsfig{file=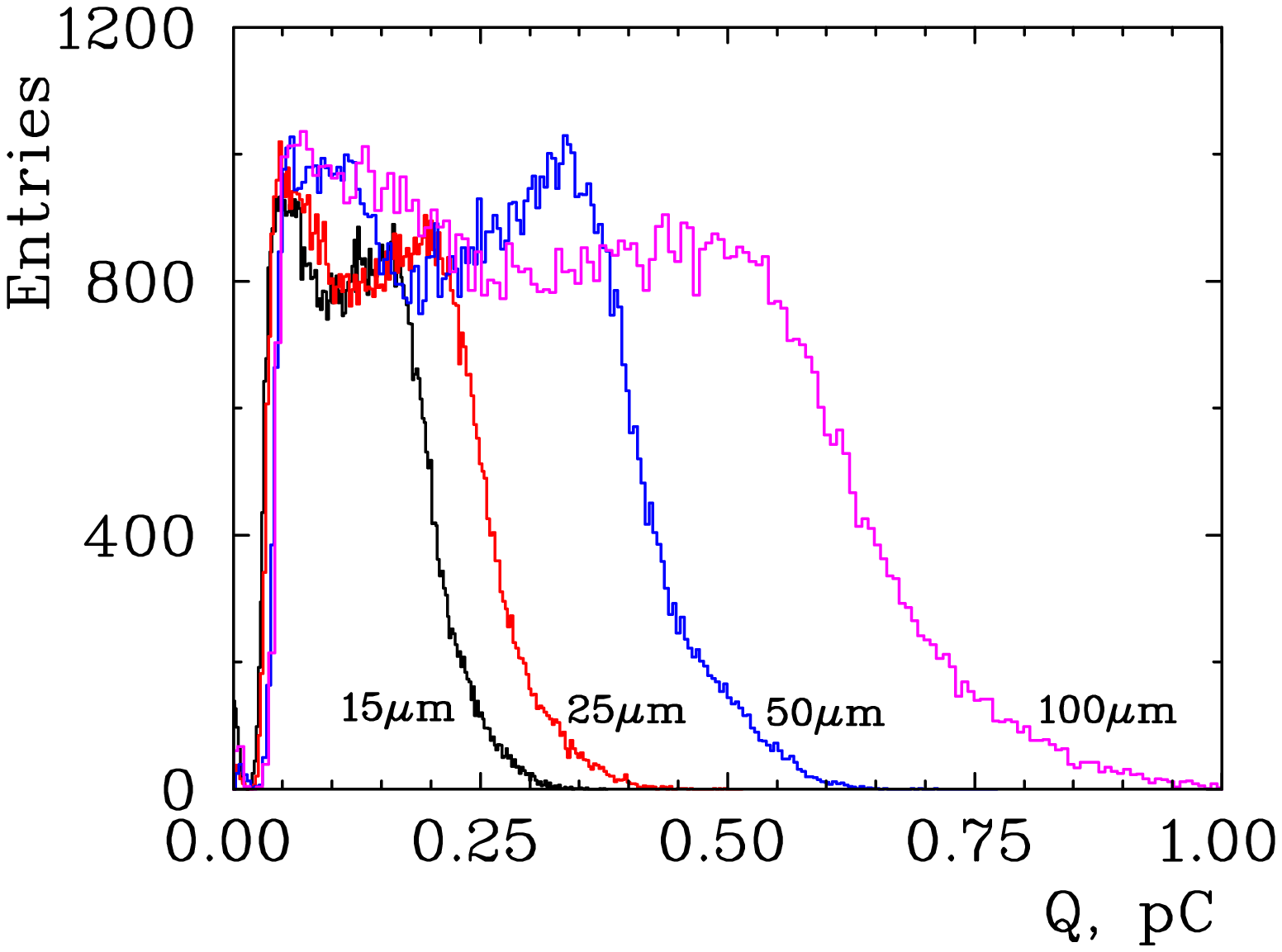,width=12cm,angle=0}
\end{center}
\caption{Measured charge spectra on  single wire chambers with 15, 25,
50   and   100~$\mu$m   diameter   anode  wires   filled   with   pure
\mbox{iso-C$_4$H$_{10}$} at  12 Torr.  All  chambers are at  800~V and
irradiated with  $^{55}$Fe x-rays. The  bigger the wire  diameter, the
higher the gas gain.}
\label{spectra_12torr}
\vspace{5mm}
\end{minipage}\hfill
\end{figure}
\indent Further  lowering of pressure results in  very poor resolution
and the  full absorption peak disappears.  At  12~Torr, charge spectra
have been taken on all four wires with 800~V applied voltage.  Spectra
were   presented    in~\cite{Davydov}   and   are    shown   here   in
fig.\ref{spectra_12torr}.  There are no indications of full absorption
peaks on any wires.  The edges of the charge distributions do indicate
gas gain on each  wire. One can see that at 12~Torr  and 800~V the gas
gains are higher on the bigger diameter wires. \\
\indent  Data taken  at  low pressure  showed  that for  a given  wire
diameter,  the resolution  of the  photoabsorption peak  degrades with
increasing  high voltage and  decreasing gas  pressure. Thin  wire has
worse  resolution  at  the  same  pressure and  applied  high  voltage
compared to thick wire. \\
\indent  In~\cite{Davydov}  an electric  field  ${E_m}$ was  introduce
where electrons gain enough energy  to ionize atoms over the mean free
path   ${\lambda_m}$,   i.e.    ${eE_m   \lambda_m  =   I_0}$,   where
${\lambda_m}=1/n\sigma$~\cite{Engel}.    Here  ${I_0}$   is   the  gas
ionization potential, $\sigma$ is the total cross section for electron
collision with atoms, {\it n} is  the density of gas atoms, {\it e} is
the  electron charge.  Above  this electric  field strength  the first
Townsend coefficient has very weak dependence on electric field and is
defined mainly by the electron's mean free path. \\
\begin{figure}[ht]
\centering
\begin{minipage}[c]{0.8\textwidth}
\vspace{5mm}
\begin{center}
\epsfig{file=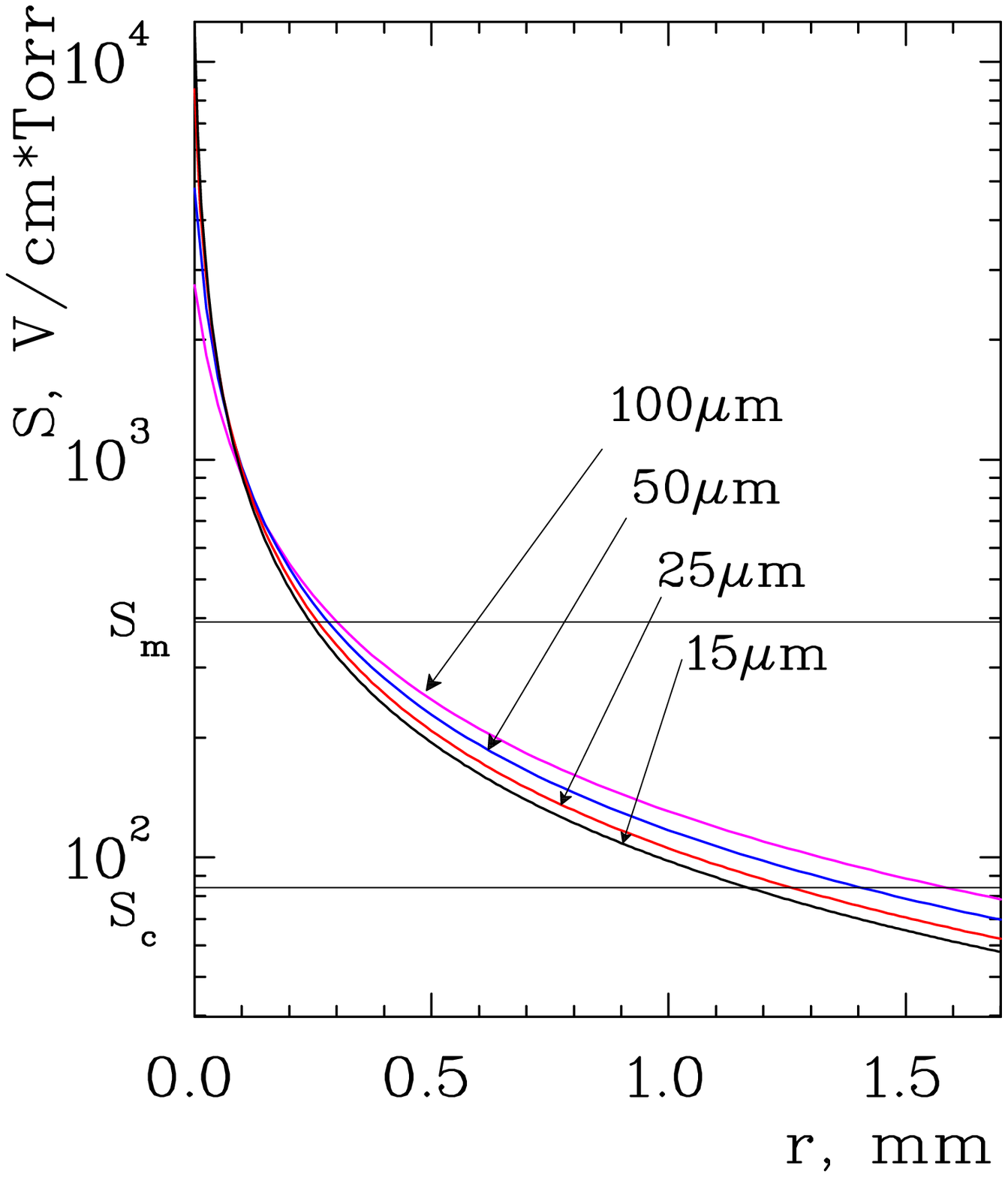,width=10cm,angle=0}
\end{center}
\caption{Reduced electric  field S on single wire chambers  with 15, 25,
50 and 100~$\mu$m diameter anode  wires as a function of distance from
the wire surfaces.  All chambers are at 800~V and pressure 12 Torr.}
\label{field_12torr}
\vspace{5mm}
\end{minipage}\hfill
\end{figure}
\indent  One   can  understand  the  gas  gain   behaviour,  shown  in
fig.\ref{spectra_12torr},  by looking  at the  reduced  electric field
strength  distributions near  the wires  and taking  into  account the
behaviour of  the first Townsend  coefficient at high  electric field,
proposed   in~\cite{Davydov}.   Figure~\ref{field_12torr}   shows  the
reduced  electric fields of  all four  tested wires  as a  function of
distance from  the wire  surfaces.  Pressure is  12~Torr and  800~V is
applied to each chamber.  The critical reduced electric field for pure
$iso-C_{4}H_{10}$,  where  the avalanche  starts,  has been  estimated
using  data  from~\cite{Baaliouamer} and  found  to  be $S_{c}  \simeq
86~V/cm\cdot Torr$. Electric  field ${E_m}$ for pure $iso-C_{4}H_{10}$
at  12~Torr  with $n=4.3\cdot  10^{17}~cm^{-3}$,  total cross  section
$\sigma  \simeq  10^{-15}~cm^{2}$~\cite{Biagi}  and 10.8~{\it  eV}  an
ionization potential~\cite{Sauli},  is estimated to  be ${E_m} \approx
4.65~kV/cm$.   These   values  of   $S_{c}$  and  the   reduced  field
$S_{m}=E_m/P  \approx  390~V/cm\cdot Torr$  are  shown  as well.   The
avalanches start farther from the surfaces of thick wires.  Everywhere
at field $S < S_{m}$  thicker wires have higher reduced electric field
that results  in a higher first Townsend  coefficient.  Beyond reduced
field $S_{m}$ the first  Townsend coefficient has very weak dependence
on electric  field and stays practically  the same for  all wires even
though the thinner wires have  much higher electric field there.  Such
a behaviour results in higher total gas gain on thicker wires. \\
\indent The  poor energy resolution in our  measurements was primarily
caused by the  small cell size (12x12~mm cross  section) of the single
wire chambers  which were  used in the  test measurements.   One would
need to use chambers with bigger cell size to improve resolution. \\
\indent  It   is  interesting  to   notice  from  the   comparison  of
fig.\ref{gain_92torr}-\ref{gain_32torr}  that  saturation  is  visible
only on 50 and 100~$\mu$m  wires at pressure 92~Torr. The highest gain
there is about 3$\cdot 10^{5}$ which corresponds to a collected charge
of about 12~pC.   The reason for this is  that photoelectrons released
by $^{55}$Fe  x-rays have much longer  range at low  pressure and many
avalanches  are   distributed  along   the  wire.   Recall,   that  at
atmospheric pressure, saturation starts when  the total charge  in the
avalanche exceeds $\approx$~1~pC.

\section{Conclusion}

\indent  We  have measured  gas  gain on  15,  25,  50 and  100~$\mu$m
diameter   wires   in    single-wire   chambers   filled   with   pure
$iso-C_{4}H_{10}$ at  pressures in the range  12-92~Torr.  Our results
clearly demonstrate that at low  pressures, gas gain becomes higher on
thick wires  than on thinner wires,  in wire chambers  having the same
geometry and  applied high voltage.  The gas gain versus  high voltage
dependence slopes  are smaller  on the smaller  diameter wires  at low
pressure.  This  is a  consequence of the  fact that at  high electric
field  strength  the first  Townsend  coefficient  is  limited by  the
electron's mean free path. \\
\indent  Bigger  wire  diameters  should  be  used  in  wire  chambers
operating at very  low gas pressures where scattering  on the wires is
not an  issue.  Specific recommendations should be  addressed for each
gas and chamber geometry.  Operating voltage (i.e. gas gain) should be
taken into  account as well because  the gas gain  versus high voltage
curves on different diameter wires have different slopes.

\end{document}